\documentclass[aps,prl,reprint,superscriptaddress]{revtex4-1}

\pdfoutput=1

\usepackage[T1]{fontenc}
\usepackage[utf8]{inputenc} 

\usepackage{lmodern}

\usepackage{tikz}

\usepackage{hyperref}
\hypersetup{colorlinks=true,allcolors=blue}
\usepackage{physics}
\usepackage{bbold}

\newcommand{\e}[1]{\mathrm{e}^{#1}}
\newcommand\heff{\hbar_\text{eff}}
\newcommand\sgn[1]{\mathrm{sgn}(#1)}

\begin{document}
\title{Chaos-Assisted Long-Range Tunneling for Quantum Simulation}
\author{Maxime Martinez}
\affiliation{Laboratoire de Physique Théorique, Université de Toulouse, CNRS, UPS, France}
\author{Olivier Giraud}
\affiliation{Université Paris-Saclay, CNRS, LPTMS, 91405, Orsay, France}
\author{Denis Ullmo}
\affiliation{Université Paris-Saclay, CNRS, LPTMS, 91405, Orsay, France}
\author{Juliette Billy}
\affiliation{Laboratoire Collisions Agrégats Réactivité, Université de Toulouse, CNRS, UPS, France}
\author{David Guéry-Odelin}
\affiliation{Laboratoire Collisions Agrégats Réactivité, Université de Toulouse, CNRS, UPS, France}
\author{Bertrand Georgeot}
\affiliation{Laboratoire de Physique Théorique, Université de Toulouse, CNRS, UPS, France}
\author{Gabriel Lemarié}
\affiliation{Laboratoire de Physique Théorique, Université de Toulouse, CNRS, UPS, France}
\affiliation{MajuLab, CNRS-UCA-SU-NUS-NTU International Joint Research Unit, Singapore}
\affiliation{Centre for Quantum Technologies, National University of Singapore, Singapore}

\begin{abstract}
We present an extension of the chaos-assisted tunneling mechanism to spatially periodic lattice systems.
We demonstrate that driving such lattice systems in an intermediate regime of modulation maps them onto tight-binding Hamiltonians with chaos-induced long-range hoppings $t_n\propto 1/n$ between sites at a distance $n$.
We provide numerical demonstration of the robustness of the results and derive an analytical prediction for the hopping term law.
Such systems can thus be used to enlarge the scope of quantum simulations to experimentally realize long-range models of condensed matter.

\end{abstract}

\maketitle

\twocolumngrid
\textit{Introduction.--} 
In recent years there has been considerable interest in the quantum simulation of more and more complex problems of solid state physics \cite{Bloch2012,Blatt2012,AspuruGuzik2012}.
In this context, lattice-based quantum simulation has become a key technique to mimic the periodicity of a crystal structure.
In such systems, dynamics is governed by two different types of processes: hopping between sites mediated by tunneling effect and interaction between particles.
While there exists several ways to implement long-range interactions \cite{Britton2012,Landig2016,Hung2016,Busche2017}, long-range hoppings have been up to now very challenging to simulate \cite{Bastidas2020,Roses2021}. 
These long-range hoppings however, have aroused great theoretical interest in condensed matter, as they are associated with important problems such as glassy physics
\cite{Sherrington1975}, many-body localization \cite{Nandkishore2017} or quantum multifractality \cite{Mirlin1996}. 
In this study we show that such long-range tunneling can be engineered in driven lattices in a moderate regime of modulation.

Temporal driving techniques are widely used in quantum simulation \cite{Eckardt2017},  as fast driving  can lead to new topological effects \cite{Lindner2011,Rechtsman2013,Goldman2014,Potirniche2017,Cooper2019} and strong driving can mimic disorder  \cite{Casati1979,Fishman1982,Graham1991,Moore1995,Casati1989,Chabe2008}.
In the intermediate regime we focus on, cold atoms in driven lattices have a classical dynamics which is neither fully chaotic (a case first explored in \cite{PhysRevE.49.70}) nor regular (corresponding to the fast driving case).
As for most real-life systems, the phase space representation of their dynamics shows coexistence of chaotic and regular zones. 
Our main result is based on the richness of the quantum tunneling in such systems, known to be \textit{chaos-assisted} \cite{Tomsovic1994,Bohigas1993,Leyvraz1996,Shudo2008,Backer2008a,
Backer2008b,Lock2010,Brodier2001,Brodier2002,Mertig2013,
Mouchet2001,Mouchet2003,Keshavamurthy2011}. 
This phenomenon is well understood between two regular islands, where it translates into large resonances of the tunneling rate between the two islands when varying a system parameter. 
It has been observed in different experimental contexts, with electromagnetic waves \cite{Dembowski2000,Hofferbert2005,Gehler2015,Shinohara2010,Backer2008a,Kim2013,Dietz2014} or cold atoms \cite{Hensinger2001,Steck2001,Steck2002, Arnal2020} (see also \cite{PhysRevE.59.2846, PhysRevLett.100.174102, xiao2013tunneling, PhysRevA.88.023810} for other related experiments). 

In this paper, we address the generalization of chaos-assisted tunneling (CAT) to \textit{mixed lattices} of regular islands embedded in a chaotic sea, obtained in a moderate regime of temporal driving.
We show that remarkably such a dynamical quantum system can be mapped onto an effective tight-binding Hamiltonian with long-range hoppings $\propto 1/n$, with $n$ the distance between sites.
Beyond the intrinsic interest of a new observable quantum chaos effect, our results open new engineering possibilities for lattice-based quantum simulations as they are highly generic, accessible for state-of-the-art experiments and species independent (in a cold atom context).

\begin{figure}
\centering
\includegraphics{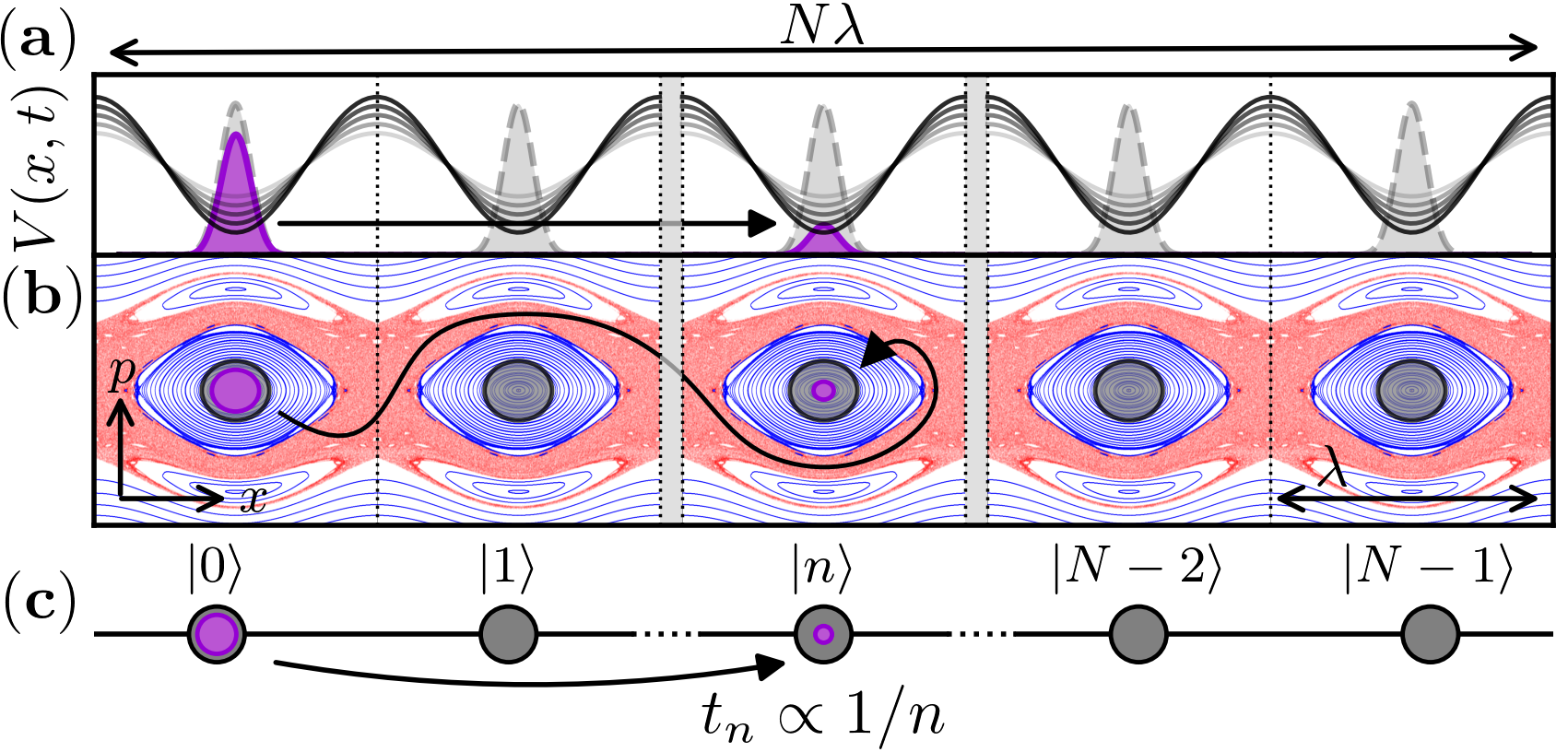}
\caption{\label{fig:principle} Three representations of CAT in a driven lattice: (a) In situ description: a wavefunction tunnels between potential wells. (b) Phase space description: the wavefunction escapes from a stable island (blue) by regular tunneling, spreads in the chaotic sea (red) and tunnels in another island. (c) Tight-binding description: the system contains $N$ sites with coupling between $i$-th and $j$-th site  proportional to $1/|i-j|$.}
\end{figure}

\textit{Model.--} 
We consider an experimental situation similar to \cite{Arnal2020}, i.e. a condensate of cold atoms in an optical lattice whose intensity is time-modulated periodically \cite{PhysRevE.49.70,Mouchet2001,Hensinger2001,Steck2001,Steck2002,PhysRevA.94.043621}. As in \cite{Arnal2020}, we assume a low density such that interactions are negligible. Using dimensionless variables \cite{NoteAdim}, the dynamics is given by the single particle Hamiltonian 
\begin{gather}
\label{eq:schrodinger}
H(x,t)=\frac{p^2}{2} - \gamma(1+\varepsilon \cos t) \cos x.
\end{gather}
$\gamma$ is the dimensionless depth of the optical lattice and $\varepsilon$ the modulation amplitude, with dimensionless time period $T=2\pi$ and spatial period $\lambda=2\pi$.
The effective Planck constant $\heff = -i [x,p]=2E_L/h\nu$ can be tuned experimentally ($\nu$ is the modulation frequency and $E_L=h^2/2 m d^2$ a lattice characteristic energy, with $d$ the lattice spacing and $m$ the atomic mass).
Beyond this model, our results are valid for almost any modulation waveform (e.g. phase modulation or kicked potentials).

\textit{Semiclassical picture.--}
The classical dynamics of this time-periodic system is best viewed through a stroboscopic phase space, using values of $(x, p)$ at each modulation period $t=j T$, $j$ integer.  For $\varepsilon=0$, the system is integrable. When $\varepsilon$ increases, chaos develops, forming a chaotic sea which surrounds regular islands of orbits centered on the stable points ($x=2n\pi$, $p=0$, $n$ an integer) of the potential wells, see Fig.~\ref{fig:principle}. 
At $\varepsilon=0$, with no chaotic sea, tunneling essentially occurs between adjacent wells, and the system can be described for deep optical lattices by an effective tight-binding Hamiltonian with nearest-neighbor hopping. Our main objective is to describe in a similar way the modulated system, a dynamical, spatially periodic lattice of $N$ regular islands indexed by $n\in [\![  0,N-1 ]\!]$, surrounded by a chaotic sea.

In a stroboscopic point of view, the quantum dynamics is described by the evolution operator $U_F$ over one period. Each eigenstate $\ket{\phi_l}$ of $U_F$ is associated with a quasi-energy $\varepsilon_l$, so that $U_F\ket{\phi_l}=\exp(-i\varepsilon_l T/\heff) \ket{\phi_l}$. 
Equivalently the Hamiltonian
$H_\text{strob}\equiv i(\heff/T)\log U_F$
gives the same stroboscopic dynamics as $U_F$ and has the same eigenstates $\ket{\phi_l}$ with \textit{energies} $\varepsilon_l$. 

\begin{figure}[htp]
\includegraphics{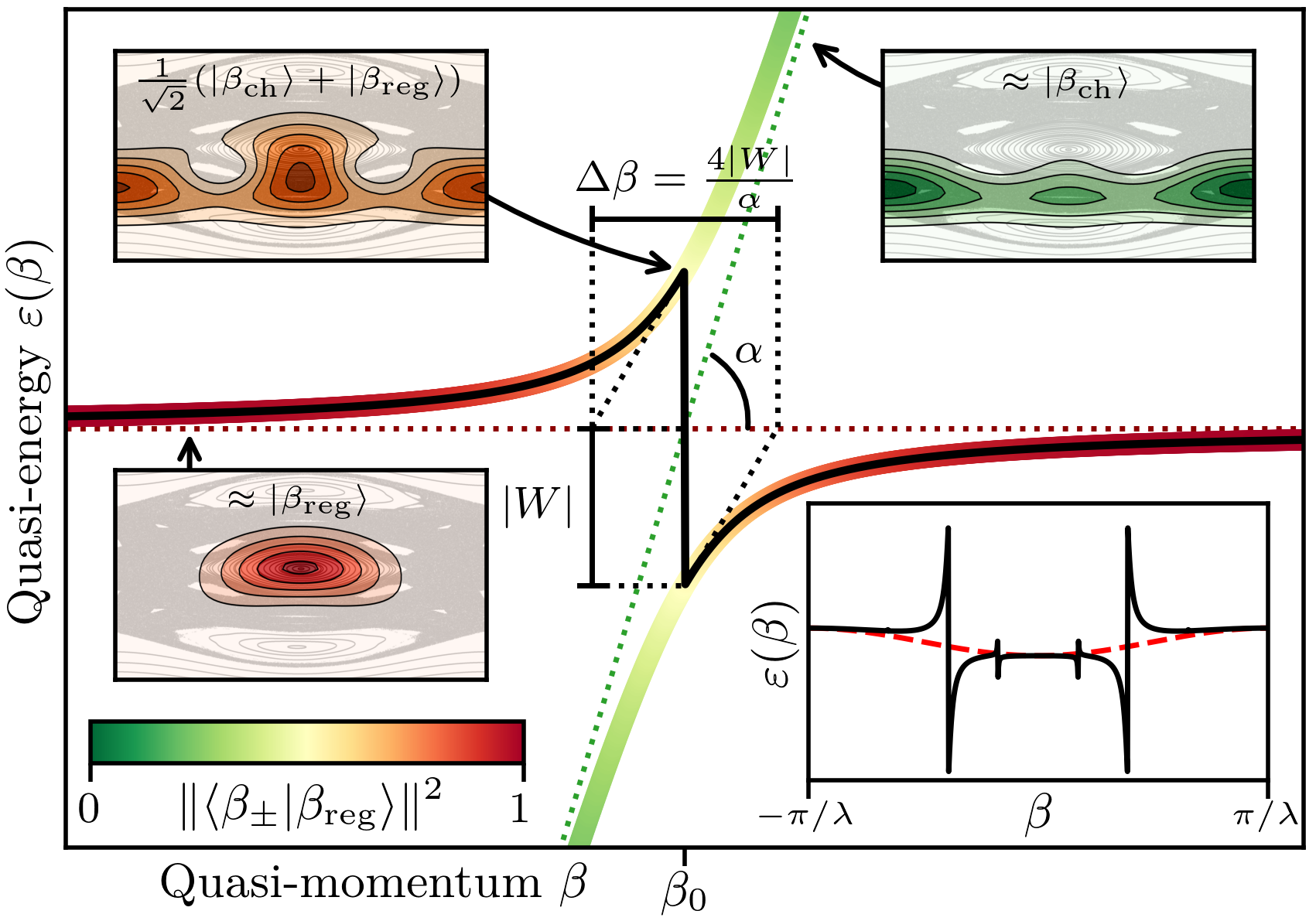}
\caption{\label{fig:crossing} In a mixed lattice as  in Fig.~\ref{fig:principle}, a CAT resonance between a regular Bloch wave $\ket{\beta_\text{reg}}$ and a chaotic one $\ket{\beta_\text{ch}}$  leads to a discontinuity in the energy band of the associated tight-binding model (solid black line).	
The main panel shows the avoided crossing characterized by $|W|$ the strength of the coupling between $\ket{\beta_\text{reg}}$ and $\ket{\beta_\text{ch}}$, $\alpha$ the slope of the energy of $\ket{\beta_\text{ch}}$, $\beta_0$ the point of equal mixing and $\Delta \beta$
the crossing width.
Near $\beta=\beta_0$, the eigenstates $\ket{\beta_\pm}$ become a mixture of $\ket{\beta_\text{reg}}$ and  $\ket{\beta_\text{ch}}$. The color code gives the intensity of the mixing (projection on $\ket{\beta_\text{reg}}$).
Husimi representations \cite{Husimi1940,Terraneo2005,Chang1986} of $\ket{\beta_\pm}$ are on top of the classical dynamics phase portrait.
Inset: Black solid line is the effective regular band and red dashed line a nearest-neighbor approximation with parameters extracted from the effective regular band at $\beta=0$ and $\beta=\pi/\lambda$ ($\heff=0.4,\gamma=0.20,\epsilon=0.15$).}
\end{figure}

In the semiclassical regime where $\heff < \mathcal A$, with $\mathcal A$ the area of a regular island, the quantum dynamics is strongly influenced by the structures of the classical phase space. Quantum eigenstates can be separated in two types \cite{Berry1977,Bohigas1993}: regular (localized on top of regular orbits) or chaotic (spread over the chaotic sea), see Fig.~\ref{fig:crossing}.

The tunnel coupling between regular states is well understood for $N=2$ regular islands surrounded by a chaotic sea (original CAT effect \cite{Bohigas1993,Tomsovic1994}). With no chaotic sea, tunneling involves only a doublet of symmetric and anti-symmetric states. In the presence of a chaotic sea, CAT is a 3-level mechanism with one of the regular states interacting resonantly with a chaotic state. This coupling leads to an energy shift and thus to a strong variation of the energy splitting giving the tunneling frequency.    
These CAT resonances, observed in a quantum system only recently \cite{Arnal2020}, occur quite erratically when varying a system parameter \cite{Tomsovic1994,Leyvraz1996}. The CAT process involves a purely quantum transport (tunneling to the chaotic sea) and a classically allowed transport (diffusion in the chaotic sea).
Thus in mixed lattices, long-range tunneling can be expected since the chaotic sea connects all the regular islands across the lattice (see Fig.~\ref{fig:principle}). 

\begin{figure}

\centering
\includegraphics[width=\linewidth]{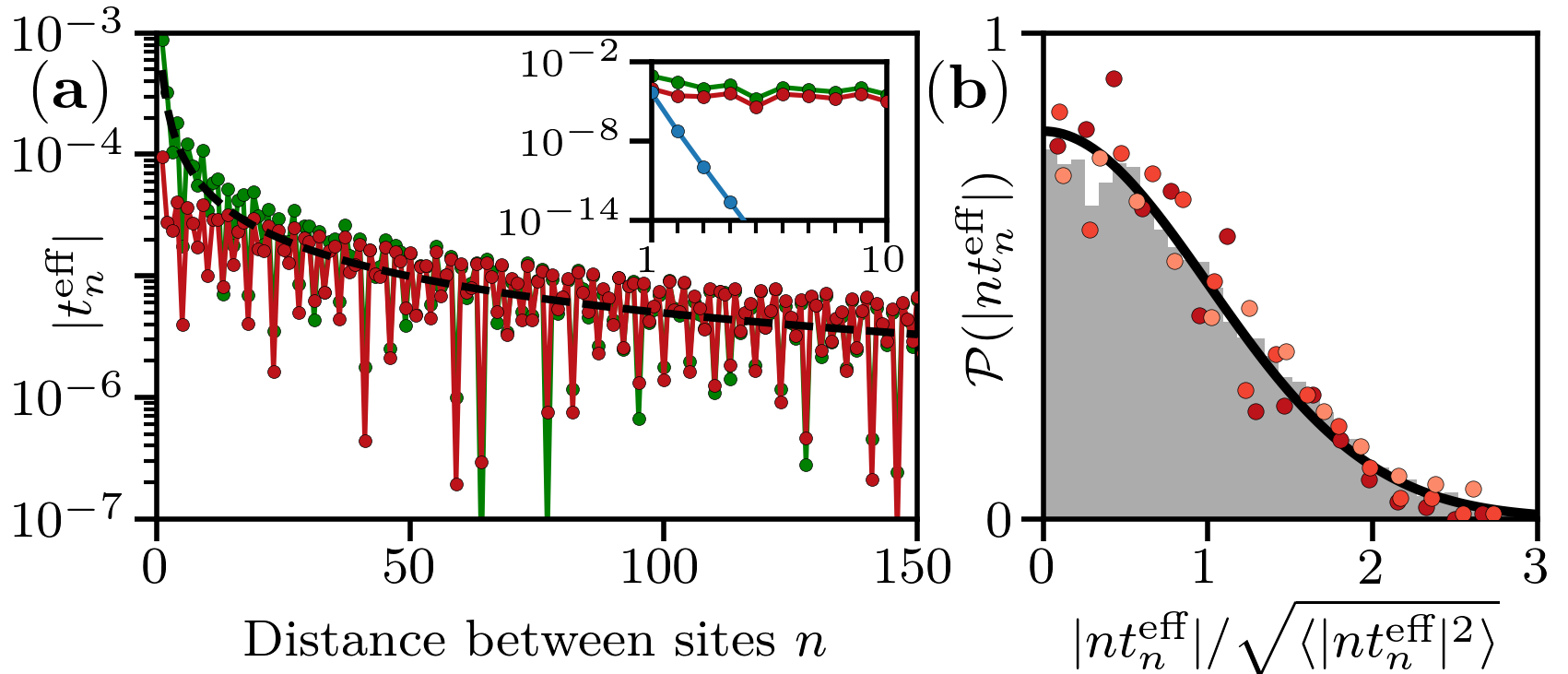}
\caption{\label{fig:coupling_analytical} (a) Effective hopping amplitude $|t_n^\text{eff}|$ vs distance between sites $n$ for $\gamma=0.2$, $\varepsilon=0.15$ and $\heff=0.4$. Red data were extracted from numerical Fourier series of the effective band structure. Green data correspond to Eq.~\eqref{eq:coupling_final} with parameters extracted from the band structure. Black solid line is the typical value of Eq.~\eqref{eq:coupling_final} (without the phase term). Inset: small-distance behavior and additional blue data for unmodulated case $\varepsilon=0$. (b) Distribution of fluctuations around the $1/n$ law for 5 parameter sets: histogram corresponds to cumulative values for $1500<n<10000$, dots are partial datasets of $500$ consecutive values of $n$, black curve is analytical prediction (see text).}
\end{figure}

\textit{Effective Hamiltonian.--} 
The existence of regular islands in the center of each cell motivates the introduction of a set of regular states $\{\ket{n_\text{reg}}\}$ (whose exact construction \cite{Backer2008a} is not crucial for our discussion) localized on these islands and forming a lattice.
For simplicity, we work in the regime $\heff\lesssim \mathcal{A}$ with only one regular state per island. 
In contrast with regular lattices, where tunneling couples only neighboring sites, there exists an \textit{indirect} coupling between distant islands of the modulated lattice mediated by the delocalized chaotic states. As in the original CAT scenario, we can expect the overlap with the chaotic sea to remain small at any time.
This motivates to capture the physics of tunneling in our system through an effective Hamiltonian $H_\text{eff}$, acting only in the regular subspace but generating the same dynamics as $H_\text{strob}$ in this subspace \cite{Feshbach1958,Feshbach1962,Lowdin1962}. Thus, the effective quantum propagator $(E-H_\text{eff})^{-1}$ (Green's function at energy $E$) should be equal to the exact one projected onto the regular subspace $ P_\text{reg} (E-H_\text{strob})^{-1}  P_\text{reg}$. The main consequence of this relation is that the effective spectrum of $H_\text{eff}$ should be included in that of $H_\text{strob}$ (see below). Thus, in the effective picture, coupling with chaotic states simply translates in a shift of the energy of each regular Bloch state $\ket{\beta_\text{reg}}=\frac{1}{\sqrt{N}}\sum_n \exp(i\beta \lambda n)\ket{n_\text{reg}}$ (with $\beta$ an integer multiple of $2 \pi/\lambda N$).
The resulting dressed regular band $ \varepsilon_\text{reg}^\text{eff}(\beta)$ then gives access to the effective tunneling coupling $t_n^\text{eff}\equiv \mel{(m+n)_\text{reg}}{H_\text{eff}}{m_\text{reg}}$ through the Fourier transform in quasi-momentum
\begin{gather}
t_n^\text{eff} = \frac{1}{N}\sum_{\beta} \varepsilon_\text{reg}^\text{eff}(\beta)\exp(i\beta \lambda n).
\label{eq:tneff}
\end{gather}
The simplest way to determine the effective spectrum is to choose the $N$ most relevant energies in the full exact spectrum.
The natural choice is to select energies of eigenstates with the largest projection on the regular subspace. In {\it mixed lattices}, this gives systematic discontinuities in the effective band, due to accidental degeneracies between a regular $\ket{\beta_\text{reg}}$ and a chaotic state $\ket{\beta_\text{ch}}$.
Close to such avoided crossings, the branch giving the effective regular energy changes, giving a sharp discontinuity of $\varepsilon_\text{reg}^\text{eff}(\beta)$ (see Fig.~\ref{fig:crossing}).
These discontinuities cause, from the Fourier transform in \eqref{eq:tneff}, a long-range decay of the effective coupling term $t_n^\text{eff}\sim 1/n$ (see Fig.~\ref{fig:coupling_analytical}).

The two main features of these resonances come from the mixed nature of the system (see Fig.~\ref{fig:crossing}):
(i) They are sharp because the local slope $\alpha$ of the crossing state is large, chaotic states being delocalized, thus sensitive to boundary conditions. (ii)
Their heights $2|W|$ are larger than the regular band width (nearest-neighbor hopping amplitudes in the regular case $\varepsilon=0$).

\begin{figure}
	\includegraphics{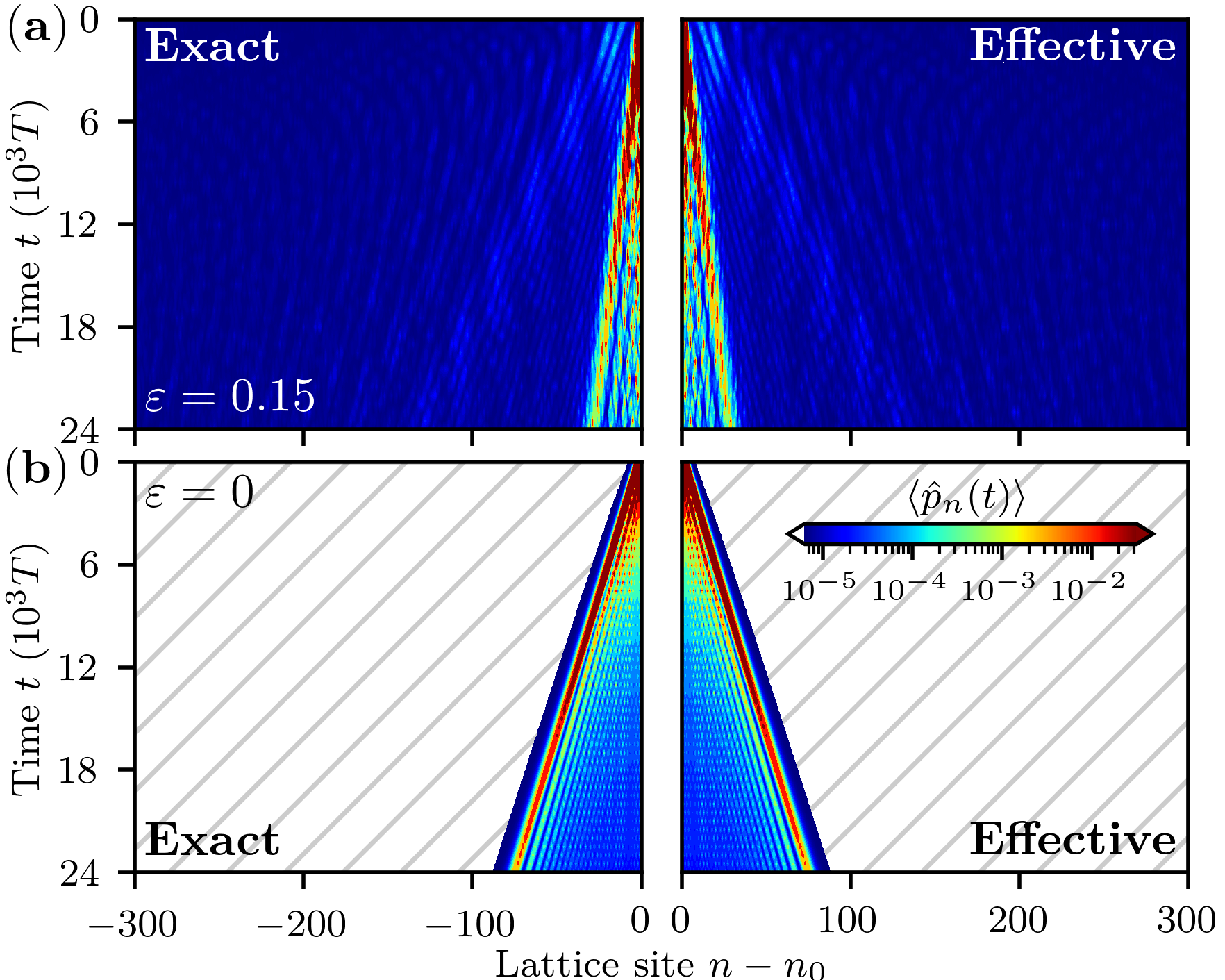}
	\caption{\label{fig:dynamics} Dynamics of a wavepacket, initially located on a
single regular island/site $n_0$. Color plot of the time evolution of the spatial probability distribution, with $\gamma=0.2$, $\heff=0.4$ and $\varepsilon=0.15$ for the modulated lattice (a) or for the unmodulated lattice, i.e. $ \varepsilon=0$ (regular case) (b).
		The exact dynamics (left) is compared with the corresponding effective description (right). Note that the system is symmetric through $n-n_0 \rightarrow n_0-n$. }
\end{figure}

\textit{Numerical simulations.--} To test the accuracy of this effective tight-binding picture, we compare the exact stroboscopic dynamics with the one given by the effective Hamiltonian, considering a wave packet initially localized on a single regular island of the modulated lattice.
(see \cite{SuppMat} for details).
As concerns the exact dynamics, the initial condition was chosen to be a localised (Wannier) state of the undriven lattice ($\varepsilon=0$), in the regular island $n_0=(N-1)/2$, $N$ being odd. 
We also used the localized states $\ket{n_\text{reg}}$ to estimate the projection of the wavefunction on the chaotic layer through $P_\text{ch}\equiv 1- P_\text{reg}$. 
The effective dynamics was studied by propagation of a state initially located at the site $n_0$ with the effective Hamiltonian.
In both simulations, we used a local observable $\hat{p}_n$ which probes the probability at each site, defined as $\hat{p}_n\equiv \ketbra{n}{n}$ in the effective system and $\hat{p}_n\equiv \int_{n\lambda}^{(n+1)\lambda} \ketbra{x}{x} \dd{x}$ in the exact one, (this choice ensures that $\sum_n \hat{p}_n=\mathbb{1}$ in both cases) and a global observable $\widehat{\Delta n^2}=\sum_n (n-n_0)^2 \hat{p}_n$ to estimate the spreading of the wave function.

\begin{figure}
		\includegraphics{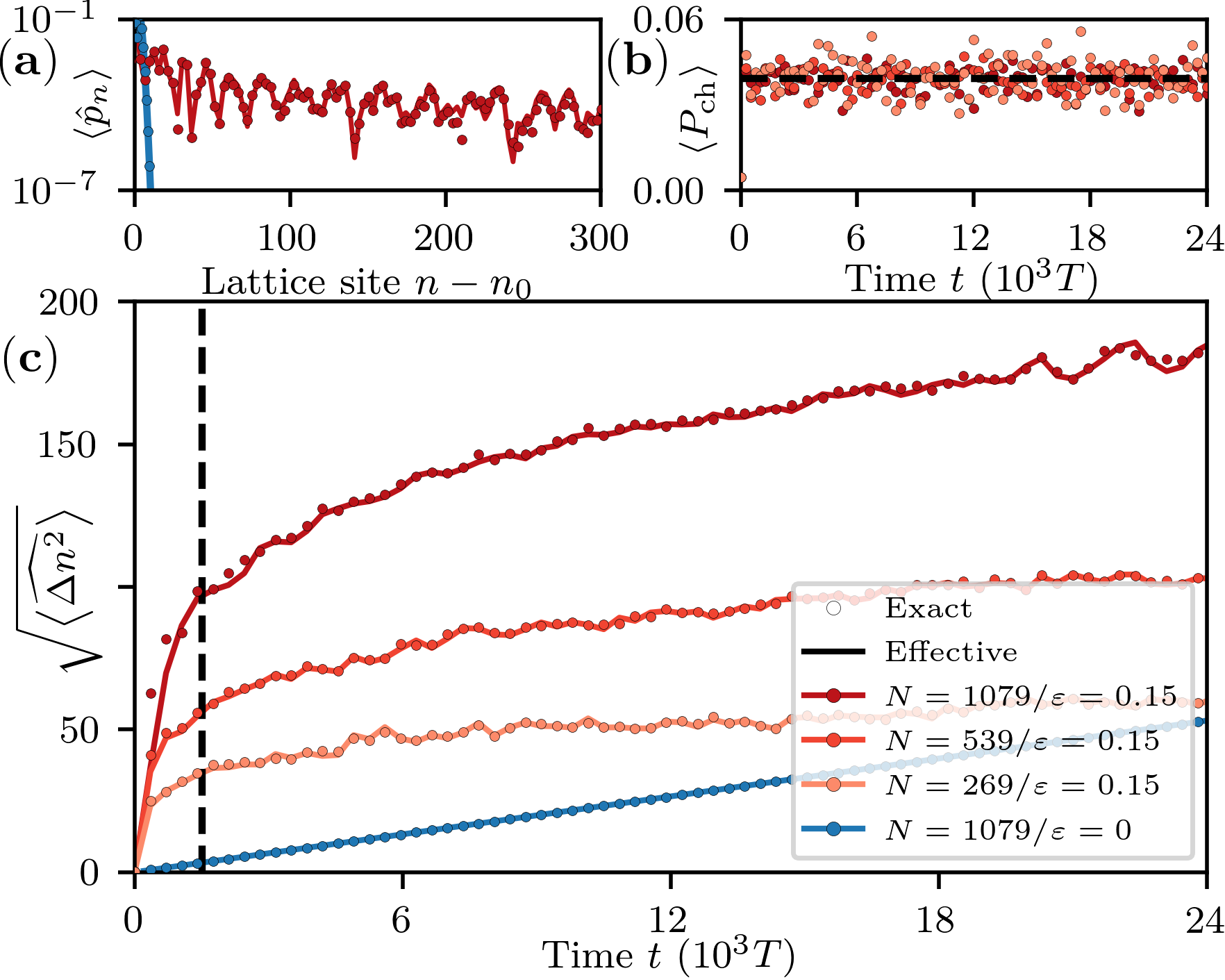}
		\caption{\label{fig:dynamics-new}Characterization of the dynamics of a wavepacket initially located on a single regular island/site $n_0$ (corresponding to Fig.~\ref{fig:dynamics}). (a) Spatial probability distribution of the wave-packet after $t=1500T$. (b) Overlap of the wavefunction with the chaotic sea vs time (see text).
				(c) Standard deviation of the spatial distribution vs time.
		Symbols are for the exact dynamics and solid lines for the effective dynamics. 
		Red data correspond to modulated lattices with different sizes, and blue data correspond to the unmodulated lattice (regular case). 
		 }
\end{figure}

We simulated different system sizes up to $N = 1079$ with periodic boundary conditions and found a very good agreement between the two approaches (see Figs.~\ref{fig:dynamics} and \ref{fig:dynamics-new} and \cite{SuppMat} for additional results).
In the modulated case, there is a fast and long-range spreading of the wavefunction (Fig.~\ref{fig:dynamics}a), in particular long tails of the spatial distribution (Fig.~\ref{fig:dynamics-new}a), that is responsible for the tremendous growth of the standard deviation (Fig.~\ref{fig:dynamics-new}c).
 The standard deviation saturates with a strong finite size effect, an additional signature of the long-range tunneling. 
In contrast, the regular $ \varepsilon=0$ case gives a slow and short-range ballistic spreading of the wavefunction with no finite-size effect (Figs.~\ref{fig:dynamics}b and \ref{fig:dynamics-new}c).

\textit{Analytical derivation of the hopping law.--} In addition to the expected long-range decay $\propto 1/n$ of the effective coupling term, numerical simulations show fluctuations around this algebraic law (see Fig.~\ref{fig:coupling_analytical}). We can explain them with a simple model: for each of the $N_\text{res}$ resonances in the effective band, we apply a two-level model with only three parameters (see Fig.~\ref{fig:crossing}): the slope $\alpha=\mathrm{d}\varepsilon_\text{ch}/\mathrm{d}\beta$ of the energy of a chaotic state with $\beta$, the coupling intensity $W$ between chaotic and regular states and the position $\beta_0$ of the crossing in the spectrum.
Using the linearity of Eq.~\eqref{eq:tneff} and assuming sharp resonances ($\Delta \beta \ll 2\pi/\lambda$), the asymptotic behavior of $t_n^\text{eff}$ is (see \cite{SuppMat})
\begin{gather}
\label{eq:coupling_final}
t_n^\text{eff}\approx \frac{i}{\pi n}\sum_{\text{resonances}} \sgn\alpha|W|\e{i n\beta_0\lambda}.
\end{gather}
This model is in very good agreement with numerical data (see Fig.~\ref{fig:coupling_analytical} and \cite{SuppMat}) and shows that the relevant time scale of the tunneling dynamics is $\heff/|W|$.
The phase term $\e{i n\beta_0\lambda}$, which depends on the position of the resonances in the effective band, gives the observed fluctuations of hopping amplitudes around the algebraic law.

Since the $W$’s of the $N_\text{res}$ resonances are associated with tunnel coupling to chaotic
states, Random Matrix Theory suggests that they can be described as
independent Gaussian variables with a fixed variance $w^2$.  In the
same spirit, as soon as $n$ is large enough the phases $n\beta_0 \lambda\;
\mod[2\pi]$ can be considered random.  Using the known results about sums of complex numbers
with  Gaussian amplitudes  and  random phases \cite{Beckmann1962}, Eq.~\eqref{eq:coupling_final}
leads to a simple statistical model for the couplings, with
$|t_n^\text{eff}| \equiv \mathcal{W}/n$ with $\mathcal{W}$ a Gaussian random
variable of variance  $N_\text{res}w^2$.  We stress that this implies the
distribution of $n | t_n^\text{eff} | $ is universal. Fig.~\ref{fig:coupling_analytical}b shows the validity of this approach.

\textit{Discussion.--} 
The theoretical results presented above rely on the effective Hamiltonian picture. It is thus important to assess its validity in our context.
The exact tunneling dynamics between two sites can be written ${\mel{(n+m)_\text{reg}}{U_F}{m_\text{reg}}=\frac{1}{N}\sum_\beta \e{i\beta \lambda n} \mel{\beta_\text{reg}}{U_F}{\beta_\text{reg}}}$. In the effective approach $\mel{\beta_\text{reg}}{U_F}{\beta_\text{reg}}$ is $\exp(-i\varepsilon_\text{reg}^\text{eff}(\beta) t /\heff)$, which does not take into account the Rabi oscillations of each regular Bloch wave $\ket{\beta_\text{reg}}$ with the chaotic sea $\ket{\beta_\text{ch}}$, whose amplitude is given in a two-level approximation by $W/\sqrt{W^2+\Delta^2}$ and whose period is $\pi \heff /\sqrt{W^2+\Delta^2}$ ($\Delta$ being the energy difference with the chaotic state involved).
The effective picture is thus valid since (i) the sharpness of resonances guarantees that the total part of the system that is delocalized in the chaotic sea is small at any time (the oscillation amplitude being large only close to the resonances), and (ii) the slowest Rabi oscillation is from \eqref{eq:coupling_final} always faster than the induced tunneling process
($\heff/| t_n^\text{eff} | \ge \pi\heff/W$).
This is confirmed by Fig.~\ref{fig:dynamics-new}b: the projection of the system on the chaotic sea displays fast and weak oscillations around a very low value.



\textit{Experiments.--} The regime of parameters we considered is experimentally relevant (lattice of depth $5 E_L$ and $\nu\approx40~$kHz as it was achieved recently~\cite{Arnal2020}).
Two complementary approaches could provide direct experimental signatures of long range tunneling: the \emph{in-situ} imaging of the cloud shape dynamics and the use of Bloch oscillations generalized to amplitude modulated lattices~\cite{PhysRevLett.120.213201, PhysRevLett.122.010402} that provides a direct spectrometry of the band from which long range properties could be inferred \cite{Stockhofe2015,Dreisow2011,Wang2010}.

\textit{Conclusion.--} In this letter we generalized the original chaos-assisted tunneling mechanism between two wells to spatially periodic lattice systems.
We demonstrated that in an intermediate regime of temporal driving, the system dynamics could be mapped onto a tight-binding Hamiltonian with long-range hopping. This is a direct consequence of the existence of sharp tunneling resonances in the band structure~\cite{Arnal2020}. These properties are a generic and robust feature of driven lattices whose classical dynamics is mixed. This effect could thus be observed in many different experimental situations.
 
Our study opens new possibilities for quantum simulation. Firstly the versatility of mixed systems allows to engineer more complex Hamiltonians such as chain of dimers with long range hoppings (with two islands per cell, see~\cite{Arnal2020}), i.e.~an extended Su-Schrieffer-Heeger model, that features non-trivial topological properties \cite{Perez2018}. Secondly, long-range hoppings in disordered lattices can generically induce non-ergodic delocalized states with multifractal properties (like in power-law random banded matrices \cite{Mirlin1996}).
Hence by adding disorder in the system, this framework should provide a way to experimentally observe quantum multifractality \cite{RevModPhys.80.1355} which is very challenging to achieve by other means \cite{faez2009observation,richardella2010visualizing, PhysRevLett.112.234101, PhysRevE.92.032914}. Finally, a proper description of many-body effects in such systems is still missing, but we may expect that they only appear within the regular islands (where the density is high) mimicking Hubbard on-site interactions. This could allow to access experimentally many-body localization and spin glass physics, where long-range tunnelings play an important role \cite{Sherrington1975,PhysRevB.99.224203,PhysRevResearch.2.043368}.

\begin{acknowledgments}
This work was supported through the EUR grant NanoX No. ANR-17-EURE-0009 in the framework of the "Programme des Investissements d'Avenir", and research funding Grant No. ANR-17-CE30-0024. We thank Calcul en Midi-Pyrénées (CALMIP) for computational ressources and assistance. We thank O. Gauthé and B. Peaudecerf for useful discussions.
\end{acknowledgments}

\bibliographystyle{apsrev4-1.bst}
\bibliography{biblio}

\begin{thebibliography}{76}
\bibitem [{Not()}]{NoteAdim}%
  \BibitemOpen
  \href@noop {} {}\bibinfo {note} {$p=2\pi P/(M \omega d)$, $x=2 \pi X/d$ where
  $P$ and $X$ are the momentum and position along the optical lattice, $M$ is
  the atomic mass, $d$ is the lattice spacing, and the time $t$ is measured in
  unit of the modulation angular frequency $\omega$}\BibitemShut {NoStop}%
\end{thebibliography}

\end{document}


\title{ Supplemental Material for \\Chaos-Assisted Long-Range Tunneling for Quantum Simulation}
	\author{Maxime Martinez}
	\affiliation{Laboratoire de Physique Théorique, Université de Toulouse, CNRS, UPS, France}
	\author{Olivier Giraud}
	\affiliation{Université Paris-Saclay, CNRS, LPTMS, 91405, Orsay, France}
	\author{Denis Ullmo}
	\affiliation{Université Paris-Saclay, CNRS, LPTMS, 91405, Orsay, France}
	\author{Juliette Billy}
	\affiliation{Laboratoire Collisions Agrégats Réactivité, Université de Toulouse, CNRS, UPS, France}
	\author{David Guéry-Odelin}
	\affiliation{Laboratoire Collisions Agrégats Réactivité, Université de Toulouse, CNRS, UPS, France}
	\author{Bertrand Georgeot}
	\affiliation{Laboratoire de Physique Théorique, Université de Toulouse, CNRS, UPS, France}
	\author{Gabriel Lemarié}
	\affiliation{Laboratoire de Physique Théorique, Université de Toulouse, CNRS, UPS, France}
	\affiliation{MajuLab, CNRS-UCA-SU-NUS-NTU International Joint Research Unit, Singapore}
	\affiliation{Centre for Quantum Technologies, National University of Singapore, Singapore}
	
	\begin{abstract}
		In this Supplemental Material, we describe the methods used to numerically simulate the dynamical evolutions of the temporally modulated system and of the effective Hamiltonian, whose construction we explain in more details. The code we used is written in Python 3 and uses the Numpy library. It is available at \url{https://framagit.org/mmartinez/dynamics1d}. We then describe in detail the derivation of the hopping law. We also present a discussion on the relation between the classical transport in the chaotic sea and the quantum long range effect. We then give another simulation with a different set of parameters than the one in the main text showing that the effect is also visible in a more semiclassical (but still experimentally relevant) regime. We end with a discussion on the use of another quantity to characterize the spreading of the wave function.
		
	\end{abstract}
	
	\maketitle
\onecolumngrid

\subsection{Numerical methods}

\subsubsection{Exact dynamics of the periodically modulated lattice}

The system, composed of $N_c$ cells of spatial size $\lambda=2\pi$ is discretized with $N_p$ points per cell; the total basis size is thus $N_t=N_c N_p$.
We have used both a spatial $\ket{x}$ and momentum $\ket{p}$ representation.
The corresponding grids are centered around $x=0$ and $p=0$ with respective size-step:
\begin{gather}
\delta x = \frac{\lambda}{N_p} \qqtext{and} \delta p = \frac{2\pi}{\lambda}\frac{\heff}{N_c}.
\end{gather}
For the whole study, we took $ N_p = 32 $ after checking that this discretization was fine enough to faithfully represent the dynamics of the system: in particular, the total size in the $p$ direction $N_p \heff$ should be larger than the extension of the chaotic sea in momentum space.

The time propagation of a given state $\ket{\psi}$ is achieved with a symmetrized split-step method:
\begin{gather}
\ket{\psi(t+\delta t)}=U_P F U_X F^{-1} U_P \ket{\psi(t)},
\end{gather}
with
\begin{gather}
U_X = \sum_x \exp(-i\frac{V(x,t) \delta t}{\hbar}) \ketbra{x}{x}, ~U_P = \sum_p \exp(-i\frac{p^2 \delta t}{4\hbar})\ketbra{p}{p}\\
F=\frac{1}{\sqrt{N}} \sum_{x,p} \exp(-i \frac{x p}{\hbar}) \ketbra{p}{x} \qqtext{(using FFT).}
\end{gather}
The time step $\delta t=4\pi/1000$ was chosen after consistency tests.

\subsubsection{Construction of the effective Hamiltonian}

The determination of the Floquet-Bloch band is equivalent to the determination of the quasi-energy spectrum of the following Hamiltonian
\begin{gather}
H_\beta(x,t)=\frac{(p-\heff\beta)^2}{2} - \gamma(1+\varepsilon \cos t) \cos x,
\end{gather}
on a single cell $N_c=1$ (with $N_p=32$, see above), with the quasi-momentum $\beta$ taking the discrete values ${\beta_m=2\pi m/(N_c \lambda)}$, $m=0,\dots, N_c-1$.
Thus, for a system size $N_c$, we repeat $N_c$ times the following procedure (for each value of $\beta_m$):
\begin{itemize}
	\item First, we build the matrix (in $x$ reprentation) of the Floquet operator from the propagation of $N_p$ $\delta$-function states $\ket{x}$.
To do so, we use the previous split-step method over two periods of modulation $T=4\pi$ (this choice was made to be consistent with \cite{Arnal2020}, but is of no importance here).
\item Second, we diagonalize the Floquet operator and look for the eigenstate having the largest overlap with a Gaussian state centered on the regular island.
This eigenstate is associated with a complex eigenvalue $\alpha_\beta$ that gives the effective energy: 
\begin{equation}
\varepsilon_\text{eff}^\text{reg}(\beta) = - \frac{i\heff}{T} \log \alpha_\beta.
\end{equation}
\item Once we have obtained the $N_c$ values of $\varepsilon_\text{eff}^\text{reg}(\beta_m)$, we build explicitly the effective tight-binding Hamiltonian $H_\text{eff}$, whose coupling elements $t_n^\text{eff}\equiv\mel{(m+n)_\text{reg}}{H_\text{eff}}{m_\text{reg}}$ are computed from the discrete Fourier Transform:
\begin{equation}
t_n^\text{eff}=\frac{1}{N} \sum_{\beta_m} \varepsilon_\text{eff}^\text{reg}(\beta_m) \exp(i \beta_m \lambda n).
\end{equation}
\end{itemize}

\subsubsection{Dynamic evolution under the effective Hamiltonian}
The effective Hamiltonian is a tight-binding model of $N_c$ sites $\ket{n}$, with $n=0,\dots N_c-1$.
The wavefunction $\ket{\psi}$ is propagated over two periods with effective evolution propagator:
\begin{gather}
\ket{\psi(t+T)}=U_\text{eff} \ket{\psi(t)} \qqtext{with} U_\text{eff}=\exp(-i\frac{H_\text{eff} T}{\heff})\; ,
\end{gather}
obtained using a Padé approximation.

\subsubsection{Construction of the regular Wannier-states}

The Wannier states of the unperturbed lattice (with $\varepsilon=0$) provide an approximation of the regular modes $\ket{n_\text{reg}}$ of the modulated lattice discussed in the letter.
To construct them, we thus use a procedure similar to that used for the determination of the effective energy band, but using the unmodulated lattice (with $\varepsilon=0$):
For each value of $\beta_m=2\pi m/(N_c \lambda)$, $m=0,\dots N_c-1$, we diagonalize the evolution operator over two periods $T=4 \pi$ and look for the eigenstate having the largest overlap with a Gaussian state centered on the regular island.
The $p$ representation of this eigenstate gives the coefficient of the Wannier state on the partial (uncomplete) grid $p=\heff \beta + n \delta p$ of size $N_p$.
After repeating $N_c$ times this procedure, we obtain the full $p$ representation of the Wannier state (on the complete momentum basis of size $N_p N_c$).

\subsubsection{Miscellaneous}

The classical dynamics is simulated using a RK4 Runge-Kutta algorithm.
Husimi phase-space representations are computed using the procedure described e.g. in \cite{Terraneo2005}.

\subsection{Derivation of the hopping law for large system sizes}

To derive the hopping law Eq.~(3), we first decompose the effective Bloch band as a regular part $\varepsilon_0$ and a sum over all resonance terms:
\begin{gather}
\label{eq:sum_ebeta}
\varepsilon^\text{eff}_\text{reg}(\beta)=\varepsilon_0(\beta)+\sum_\text{resonances} \varepsilon_\asymp(\beta-\beta_0,W,\alpha)\,,
\end{gather}
where we assume that the evolution of the energy near each resonance is universal and only depends on three parameters: $\beta_0$ the position of the resonance, $W$ the coupling intensity between the chaotic and the regular state and $\alpha$ the slope of the energy of the involved chaotic state with $\beta$. 
More precisely we consider that the system is locally described by a two-level Hamiltonian with an avoided crossing at $\beta'=\beta-\beta_0=0$:
\begin{gather}
\mqty(\varepsilon_\text{reg}(\beta') & W\\ W & \varepsilon_\text{ch}(\beta'))
\end{gather}
with $\varepsilon_\text{reg}(\beta')=0$ (since it is taken into account by $\varepsilon_0$ in Eq.~\eqref{eq:sum_ebeta}) and $\varepsilon_\text{ch}(\beta')=\alpha\beta'$. The corresponding eigenstates $\vert \beta_\pm\rangle$ and eigenenergies $\varepsilon_\pm(\beta')$ follow:
\begin{gather}
\label{eq:shiftenergy}
\varepsilon_\pm(\beta')= \frac{\varepsilon_\text{reg}+\varepsilon_\text{ch}}{2} \pm \sqrt{\Delta^2 + W^2}\qqtext{and} \ket{\beta_\pm}=\begin{cases}
\cos \theta \ket{\beta_\text{reg}} + \sin \theta \ket{\beta_\text{ch}}\\
-\sin \theta \ket{\beta_\text{reg}} + \cos \theta \ket{\beta_\text{ch}}
\end{cases},
\end{gather}
with $\Delta=(\varepsilon_\text{reg}-\varepsilon_\text{ch})/2$ and $\theta \in [0,\pi/2]$ verifying $\tan 2\theta=|W|/\Delta$. 
The prescription for the effective spectrum construction is to select the energy associated with the eigenstate having the largest projection on the regular subspace. We thus get:
\begin{gather}
\varepsilon_\asymp(\beta',W,\alpha) = \frac{\alpha}{2}\qty(\beta'-\sgn{\beta'} \sqrt{\beta'^2+\qty(\frac{2|W|}{\alpha})^2}).
\end{gather}

Taking the Fourier transform, we then have
\begin{gather}
t_n = t_n^0+\sum_\text{resonances} t_n^\asymp(\beta_0,W,\alpha) \qqtext{with}
t_n^\asymp(\beta_0,W,\alpha)=\frac{\lambda}{2\pi}\int_{-\pi/\lambda}^{\pi/\lambda} \varepsilon_\asymp(\beta-\beta_0,W,\alpha) \e{-i n\beta \lambda} \dd{\beta}.
\end{gather}

We now assume that $\varepsilon_\asymp(\beta-\beta_0,W,\alpha)$ is peaked around $\beta_0$ and that $\beta_0$ is sufficiently far from the edge of the boundary of the Brillouin zone, so that
\begin{gather}
t_n^\asymp(\beta_0,W,\alpha)\approx \e{in\beta_0 \lambda} \frac{\lambda}{2\pi}\int_{-\pi/\lambda}^{\pi/\lambda} \varepsilon_\asymp(\beta,W,\alpha) \e{-i n\beta \lambda} \dd{\beta}.
\end{gather}

The latter expression can be evaluated for large $n$ values.
We introduce $x=\beta \lambda$ and $\eta=2 \lambda|W|/\alpha=\lambda \Delta\beta/2$, it reads
\begin{align}
t_n^\asymp=&\frac{\e{in\beta_0 \lambda} \alpha}{4\pi \lambda} \times \underbrace{\int_{-\pi}^{\pi} \qty(x-\sgn{x}\sqrt{x^2+\eta^2})\e{-inx} \dd{x}}_{I^*},
\end{align}
we split the integral $I$ (taking complex conjugation) in two parts, the first one gives
\begin{gather}
\int_{-\pi}^{\pi} x\e{inx} \dd{x}  = \frac{2i\pi}{n} (-1)^{n+1}.
\end{gather}
The second part can be rewritten
\begin{gather}
\int_{0}^{\pi} \sgn{x}\sqrt{x^2+\eta^2}\e{inx} \dd{x} - \text{c.c.},
\end{gather}
we then deform the contour of integration $0\rightarrow \pi$ to a complex circuit $0 \rightarrow iT \rightarrow iT+\pi \rightarrow \pi$ with $T$ some large real number. Using Watson's formula, the first part gives (setting $x=iy$)
\begin{gather}
i\int_{0}^{T} \sqrt{\eta^2-y^2}\e{-ny} \dd{y}\sim \frac{i |\eta|}{n}.
\end{gather}
The second part is negligible for $T$ large enough (setting $x=y+iT$):
\begin{gather}
\e{-nT} \int_0^\pi \sqrt{(y+iT)^2+\eta^2} \e{-iny} \dd{y} \rightarrow 0.
\end{gather}
Using Watson's formula and assuming $\Delta \beta \ll \frac{2\pi}{\lambda}$ so that $(\eta/\lambda)^2 \ll 1$, the third part (setting $x=\pi + iy$) gives:
\begin{align}
i(-1)^{n+1}\int_0^T \sqrt{(\pi+iy)^2+\eta^2} &\e{-ny} \dd{y} \sim  \frac{i\pi}{n}(-1)^{n+1}.
\end{align}
Putting all terms together (taking care of complex conjugation) we end up with
\begin{gather}
t_n^\asymp\approx \frac{\e{in\beta_0 \lambda} \alpha}{4\pi \lambda} \qty(\frac{2i\pi}{n}(-1)^{n+1}-\frac{2i \eta}{n}-\frac{2i\pi}{n}(-1)^{n+1})^* = \frac{\e{in\beta_0 \lambda} \alpha}{4\pi \lambda} \times \frac{i 4 \lambda |W|}{|\alpha|} =\frac{i}{\pi n}\sgn\alpha|W|\e{i n\beta_0\lambda}.
\end{gather}

We finally assume that $t_n^0$ is negligible for large $n$ values (because it decays exponentially), so that
\begin{gather}
t_n \approx \frac{i}{\pi n}\sum_{\text{resonances}} \sgn\alpha|W|\e{i n\beta_0\lambda}.
\end{gather}

\subsection{Classical dynamics and quantum long-range properties}

\begin{figure}[!h]
\centering
\includegraphics[width=0.85\linewidth]{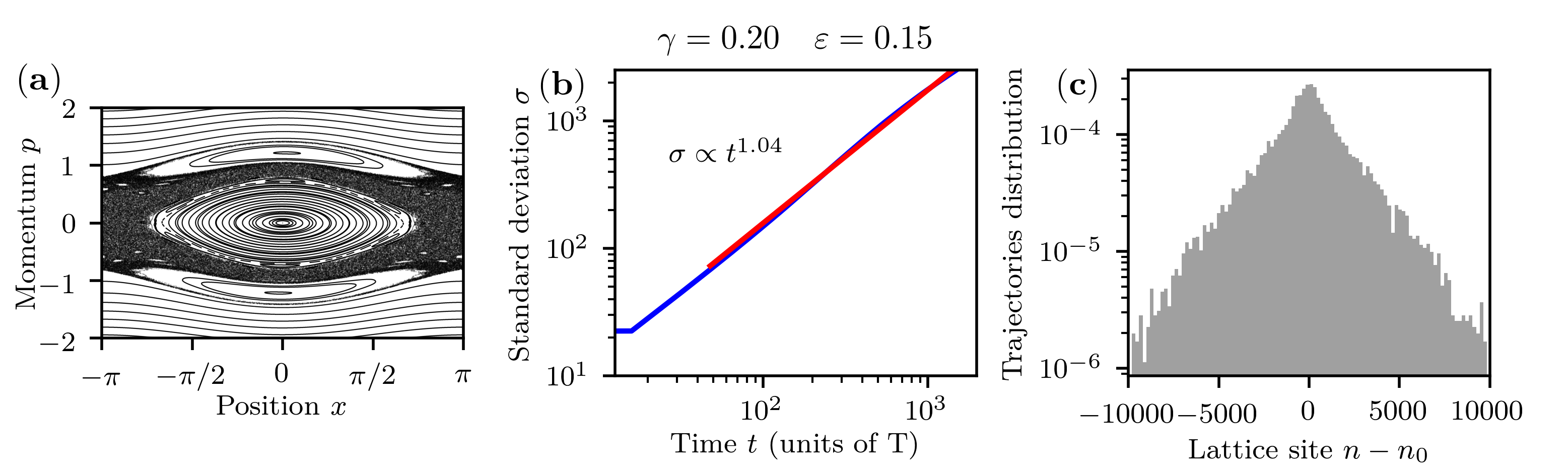}
\includegraphics[width=0.85\linewidth]{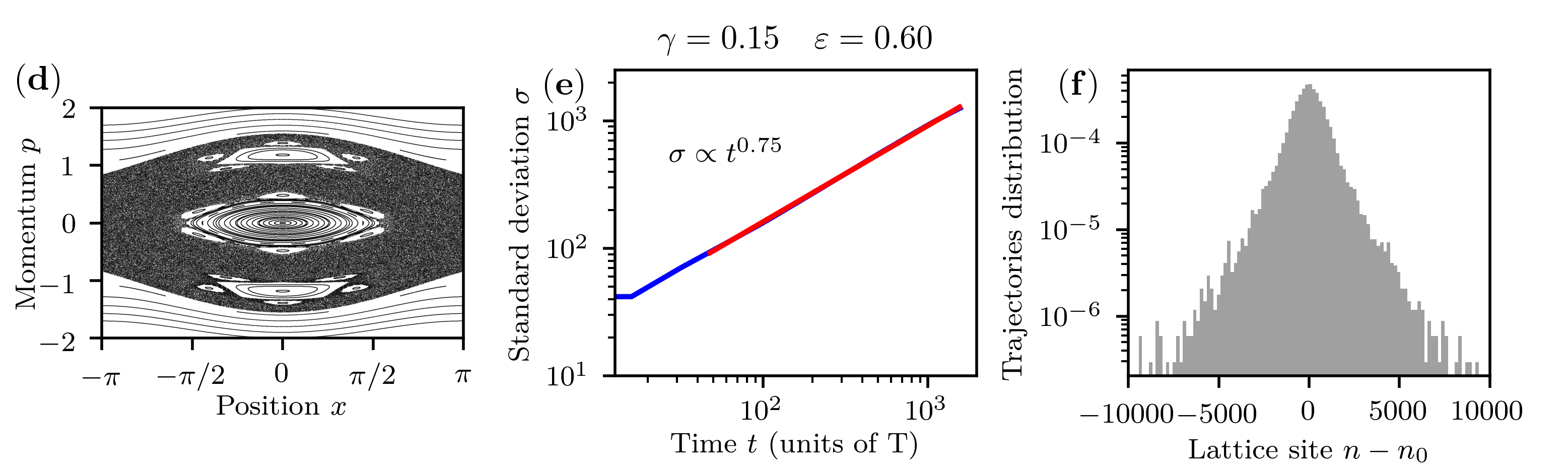}
\includegraphics[width=0.85\linewidth]{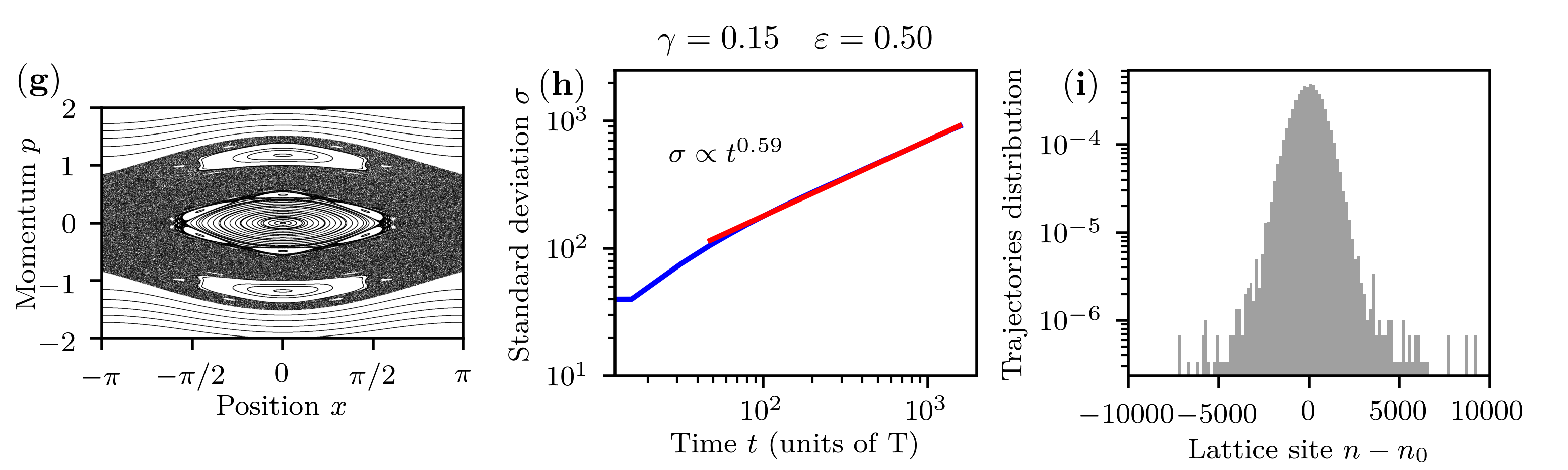}
\caption{\label{SMdynclass} Classical dynamics of Eq.~(1) of the main text, for different set of parameters ((a-c) correspond to the parameters in the main text). (a,d,g) Stroboscopic phase portraits. (b,e,h) Standard deviation of the spatial distribution of a classical wavepacket initially launched in the chaotic sea ($1799$ trajectories launched around $(x_0,p_0)=(\pi,0)$). Blue solid line for numerical data and red solid line for corresponding linear fit ($\log\sigma=a \log t + b$). The fitted value of the diffusion exponent $a$ is given in the figure.  (c,f,i) Corresponding spatial distribution at $t=10000T$.}
\end{figure}

In the main text we discuss how the long-range chaos-assisted tunneling mechanism can be qualitatively inferred from a semiclassical picture. In the other hand, its appearance at quantum scale is a direct consequence of the existence of sharp and strong tunneling resonances in the band structure, which is a fairly generic feature of system whose classical dynamics is mixed. Hence it is not clear how the corresponding classical transport properties in the chaotic sea may affect this mechanism.

\begin{figure}[htp]
\centering
\includegraphics[width=0.85\linewidth]{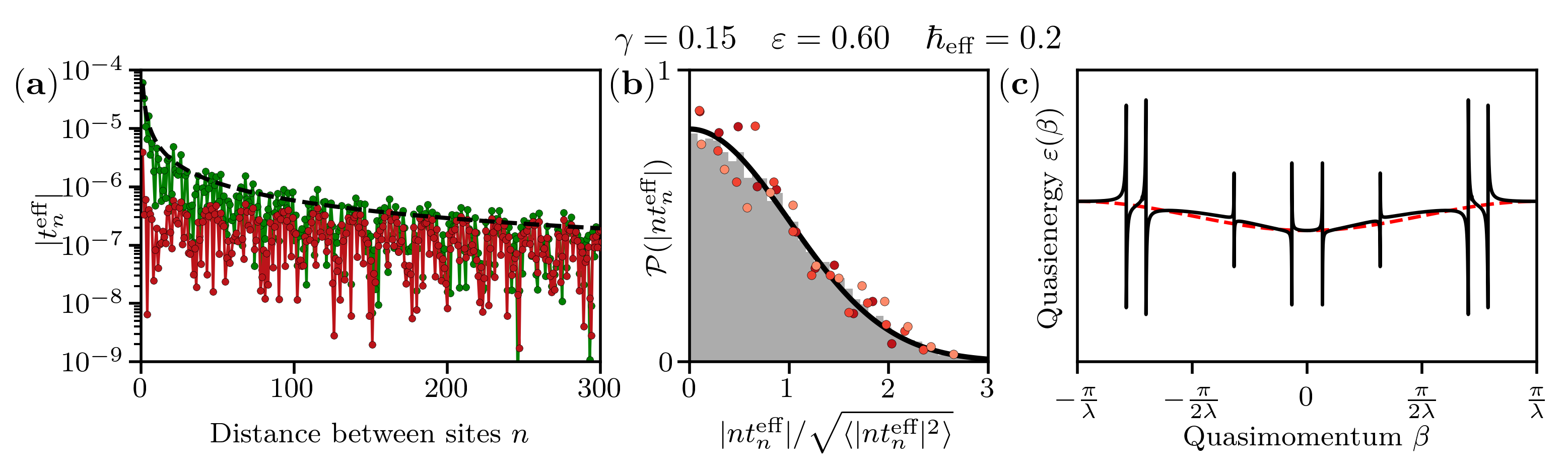}
\includegraphics[width=0.85\linewidth]{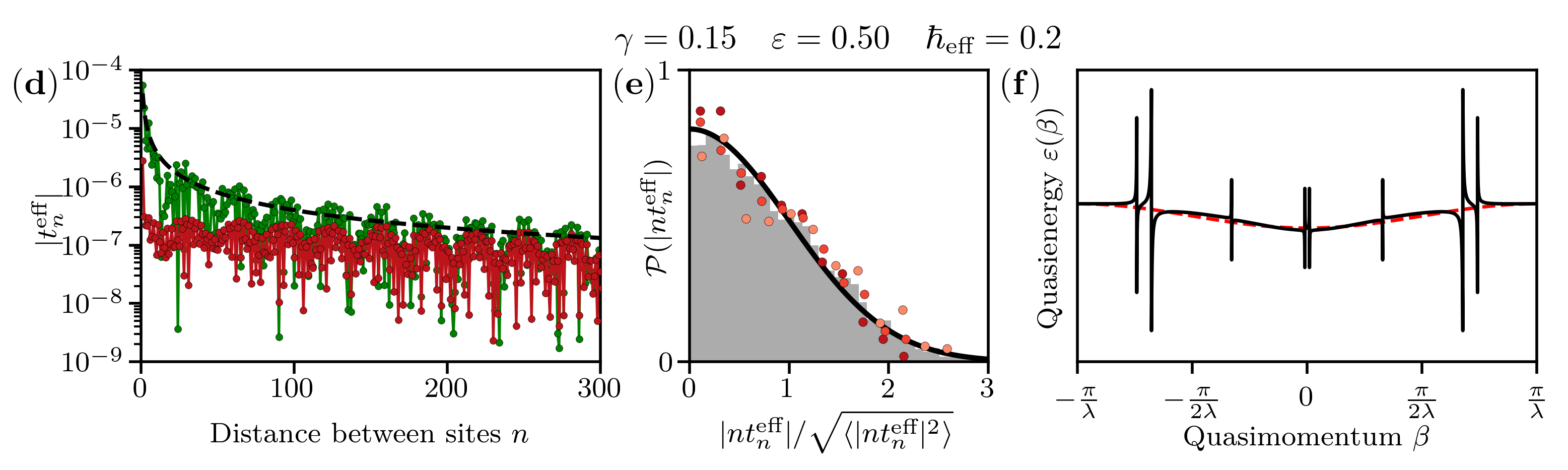}
\caption{\label{SMquantumdis} (a,d) Effective hopping amplitude $|t_n^\text{eff}|$ vs distance between sites $n$ for different set of parameters. Red data were extracted from numerical Fourier series of the effective band structure. Green data correspond to Eq.~3 of the main text with parameters extracted from the band structure. Black solid line is the typical value of Eq.~(3) of the main tex  (without the phase term). (b,e) Distribution of fluctuations around the $1/n$ for the same parameters. Histogram corresponds to cumulative values for $1500<n<10000$, dots are partial datasets of $500$ consecutive values of $n$, black curve is analytical prediction (see text).
(c,f) Quasi-energy dispersion relation, black solid line is the effective regular band and red dashed line a nearest-neighbor approximation with parameters extracted from the effective regular band at $\beta=0$ and $\beta=\pi/\lambda$.
}
\end{figure}

For instance, in the classical counterpart of the model we study, the motion in the chaotic sea is generically superdiffusive. Indeed, there exist large transporting islands (Fig.~\ref{SMdynclass} of this Supplemental Material) centered around $p=\pm 1$ (which travel at constant speed $v = \lambda/T$) around which classical trajectories can stick for a long time, accelerating the usual diffusion process in the chaotic sea.
As a consequence the value of the classical diffusion exponent depends a lot on the fine structure of the phase space, as is shown for three different sets of parameters in Fig.~\ref{SMdynclass} of this Supplemental Material. However, this does not affect the quantum long range properties  
 of the effective Hamiltonian, which remain essentially unchanged, see Fig.~\ref{SMquantumdis} of this Supplemental Material and Fig.~5 of the main text. Thus the quantum long range properties do not seem to depend on the details of the classical transport processes inside the chaotic sea.

\subsection{Probing a deeper semiclassical regime ($\gamma=0.15$, $\varepsilon=0.60$ and $\heff=0.2$)}

\begin{figure}[!h]
\begin{minipage}[c]{0.45\linewidth}

\centering
\includegraphics[width=\linewidth]{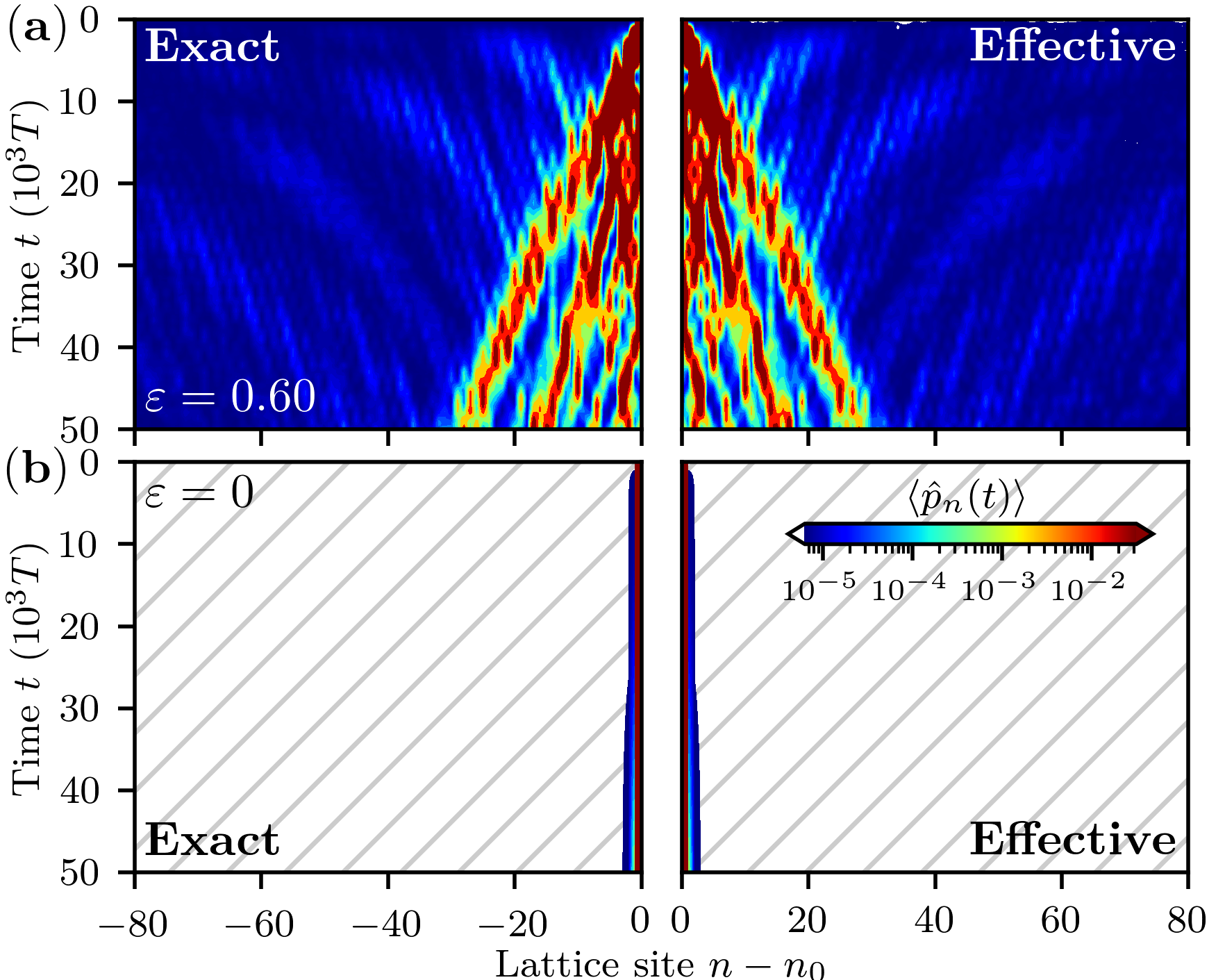}
\caption{\label{SMquantumdyn1} Dynamics of a wavepacket, initially located on a
single regular site $n_0$. Probability at each site vs time, with $\gamma=0.15$, $\heff=0.2$ and $\varepsilon=0.60$ for modulated lattice (a) or $ \varepsilon=0$ for unmodulated lattice (b).
		Exact dynamics (left) is compared with corresponding effective description (right), note that the system is symmetric through $n-n_0 \rightarrow n_0-n$. }

\end{minipage} \hfill
\begin{minipage}[c]{0.45\linewidth}
\centering
\includegraphics[width=\linewidth]{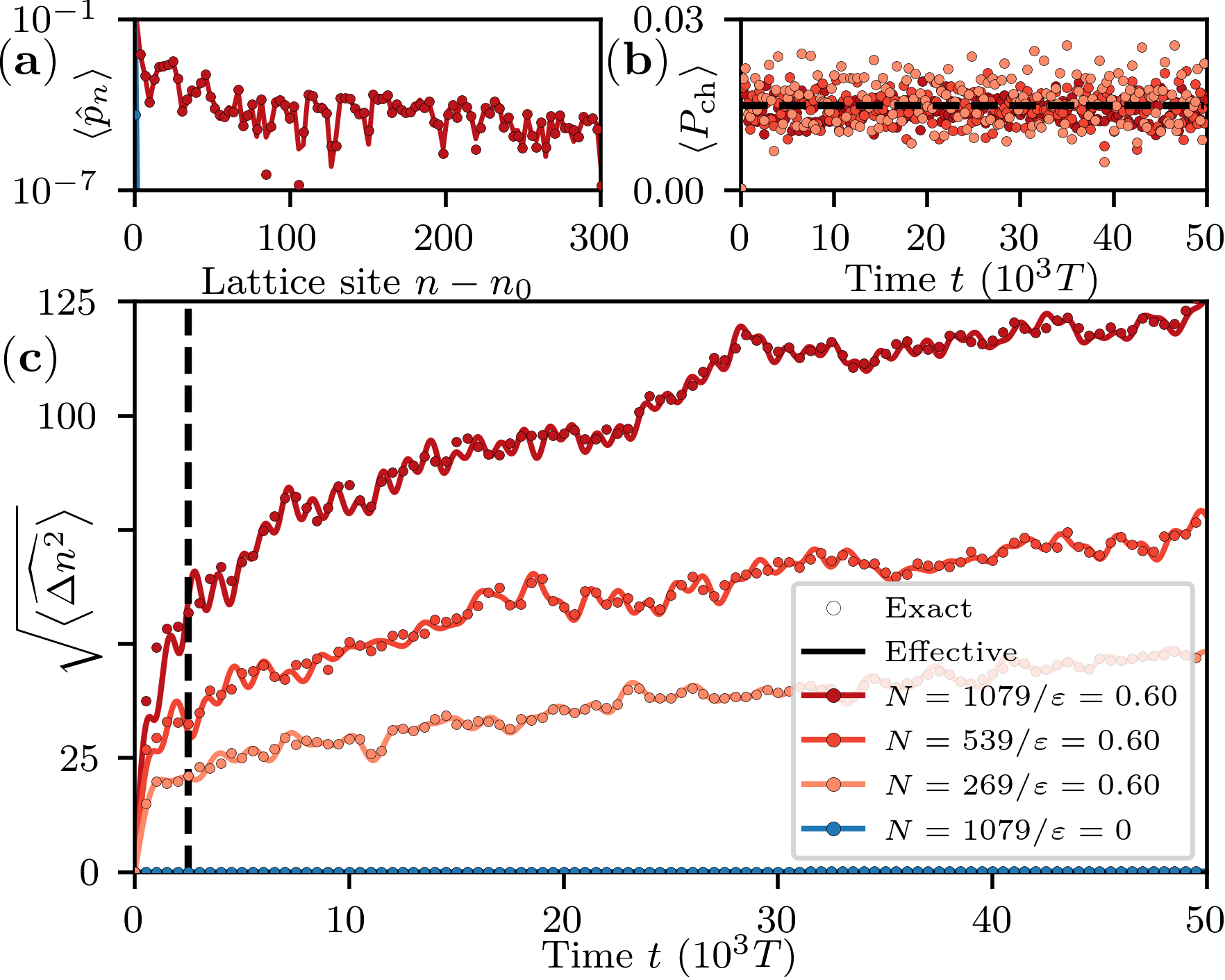}
		\caption{\label{SMquantumdyn2}Characterization of a wavepacket initially located on a single regular site $n_0$ (corresponding to Fig.~\ref{SMquantumdyn1} of this Supplemental Material). 
		Symbols for exact dynamics and solid lines for effective dynamics. 
		Red data for modulated lattices with different sizes, and blue data for unmodulated lattice. 
		 (a) Spatial probability distribution of the wavepacket after $t=2500T$. (b) Overlap of the wavefunction with the chaotic sea vs time (see text).
		 (c) Standard deviation of the spatial distribution vs time.}

\end{minipage}
\end{figure}

We here present the comparison between the exact dynamics and the effective approach of a wavepacket initially located on a single regular site, for one of the sets of parameters presented in the previous section.
The value of $\heff=0.2$ was chosen smaller than in the main text ($\heff=0.4$), to probe a more semiclassical regime (this value still remains relevant experimentally).
The effective approach appears to be as good as in the other set of parameters (compare Figs.~\ref{SMquantumdyn1} and \ref{SMquantumdyn2} of this Supplemental Material with Figs. 3 and 4 of the main text) and even better if one looks at the projection in the chaotic sea (Fig.~\ref{SMquantumdyn2}b of this Supplemental Material) that saturates around $\sim 1.5\%$ to be compared with $\sim 5\%$ in the main text (Fig.~4b of the main text).
This can be explained by the fact that in this regime the resonances are sharper.
As an additional consequence of this smaller value of $\heff$, the unmodulated dynamics is totally frozen on the time scale we chose (Fig.~\ref{SMquantumdyn1}), because the regular tunneling rate decreases exponentially with $\heff$.

\subsection{Assessing the long range properties through the inverse participation ratio}

\begin{figure}[!h]
\centering
\includegraphics[width=0.5\linewidth]{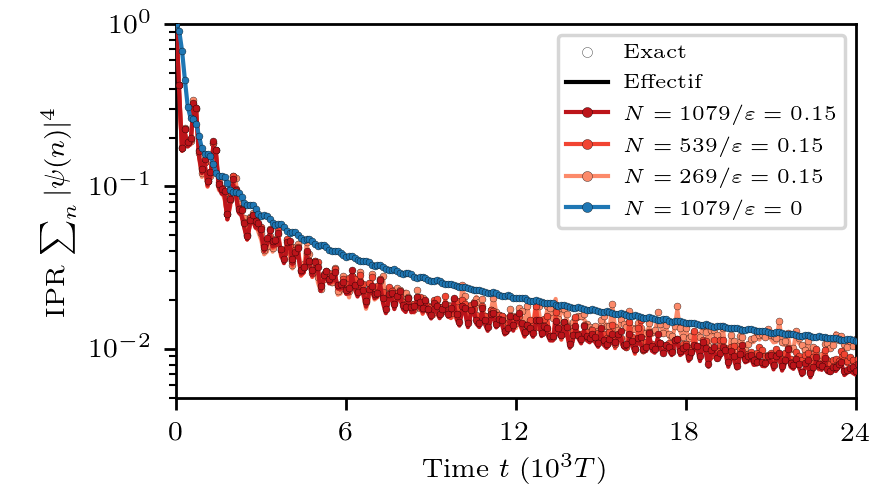}
\caption{\label{SMipr} Characterization of a wavepacket initially located on a single regular site $n_0$ (corresponding to Figs.~3 and 4 of the main text). The inverse participation ratio is plotted as a function of time. Symbols for exact dynamics and solid lines for effective dynamics. Red data for modulated lattices with different sizes, and blue data for unmodulated lattice. }
\end{figure}

In the main text we use the standard deviation of the spatial distribution of the wave packet to assess the long range properties of the system. Another popular quantity measuring the degree of localization of a wavefunction is the inverse participation ratio $\sum_n |\psi(n)|^4$.
However, quite surprisingly, the participation ratio of a wavepacket initially located on a single regular site $n_0$ appears to poorly capture the long-range properties of the dynamics (Fig.~\ref{SMipr} of this Supplemental Material).
This can be explained if we note that the inverse participation ratio is dominated by the bulk of the distribution (the large wavefunction amplitudes), while long-range properties manifest themselves in long tails of the spatial probability distribution of the wavepacket.
This also explains why the standard deviation of the spatial distribution (which magnifies the long tails of the distribution) succeeds at revealing the long-range properties of the dynamics (see Fig~4 of the main text).

\bibliographystyle{apsrev4-1.bst}
\bibliography{biblio}